\documentclass[aps,prd,twocolumn,superscriptaddress,showpacs,showkeys]{revtex4}

\usepackage{epsfig,epsf}
\usepackage{amsmath}
\usepackage{amsthm}
\usepackage{amsfonts}
\usepackage{amssymb}
\usepackage{dsfont}

\usepackage{psfrag}
\usepackage{slashed}

\def\OMIT#1{}

\newcommand{\bea}{\begin{eqnarray}}
\newcommand{\eea}{\end{eqnarray}}

\newcommand \ket [1] {|{#1}\rangle}

\newcommand{\ft}[2]{{\textstyle\frac{#1}{#2}}}
\newcommand{\bit}[1]{\mbox{\boldmath$#1$}}

\begin{document}

\title{\bf Twist-four Corrections to Parity-Violating Electron-Deuteron
Scattering}

\author{A.V. Belitsky}
\email[]{andrei.belitsky@asu.edu}
\affiliation{Department of Physics, Arizona State University, Tempe, AZ 85287-
1504, USA}
\author{Alexander Manashov}
\email[]{alexander.manashov@physik.uni-regensburg.de}
\affiliation{Institut f\"ur Theoretisch Physik, University of Regensburg, D-
93040 Regensburg, Germany}
\affiliation{Department of Theoretical Physics,  St.-Petersburg State
University, 199034, St.-Petersburg, Russia}
\author{Andreas Sch\"afer}
\email[]{andreas.schaefer@physik.uni-regensburg.de}
\affiliation{Institut f\"ur Theoretisch Physik, University of Regensburg, D-
93040 Regensburg, Germany}

\begin{abstract}
\vspace*{0.3cm}
Parity violating electron-deuteron scattering can potentially provide a clean
access to electroweak
couplings that are sensitive to physics beyond the Standard Model. However
hadronic effects can contaminate
their extraction from high-precision measurements. Power-suppressed
contributions are one of the main sources
of uncertainties along with charge-symmetry violating effects in leading-twist
parton densities. In this work we
calculate the twist-four correlation functions contributing to the left-right
polarization asymmetry making use of
nucleon multiparton light-cone wave functions.
 \end{abstract}

\maketitle

\section{Introduction}

Even after decades of experimental studies, deep inelastic scattering (DIS)
remains one of the most
powerful tools for unraveling the partonic structure of nucleons and nuclei.
DIS also allows for systematic searches for physics beyond the standard model.
Parity violation in DIS (PV-DIS) at medium energies
is particularly sensitive to effects of New Physics.
Historically, this process played an important role in verifying the Standard
Model
\cite{Prescott:1978tm,Cahn:1977uu}. Today the search for New Physics motivates a
number of
ongoing and planned experiments
\cite{Ito:2003mr,Spayde:2003nr,Anthony:2005pm,Armstrong:2005hs,Acha:2006my,Baunack:2009gy,:2009zu}.
The physics reason for this great interest is that within the standard model the
Weinberg angle
$\theta_W$ should show a highly non-trivial characteristic scale dependence,
which can be mapped out by
combining experiments at different momentum scales.
The SoLID experiment at JLab  \cite{Souder:2008zz,JLAB1} (see also \cite{JLAB2,Qwe}) will be especially
sensitive to the poorly measured weak neutral coupling constants
$C_{2 q}$ in the low-energy electroweak Lagrangian
\begin{align}
\label{PVlagrangian}
{\cal L}_{\rm\scriptscriptstyle PV}
=
\frac{G_F}{\sqrt{2}}
\bigg[&
\bar{e} \gamma^\mu \gamma_5 e
\left(
C_{1 u} \bar{u} \gamma_\mu u + C_{1 d} \bar{d} \gamma_\mu d
\right) \nonumber\\
+&
\bar{e} \gamma^\mu e
\left(
C_{2 u} \bar{u} \gamma_\mu \gamma_5 u + C_{2 d} \bar{d} \gamma_\mu
\gamma_5 d
\right)
\bigg]
\, .
\end{align}

To analyze the theoretical situation, effects of New Physics are parameterized
by $\delta C_{i \alpha}$
\footnote{Interactions mediated by $Z'$ bosons or supersymmetric partners of
observed particles
fall into the class parameterized by current-current interaction at low
energies.} according to
$C_{1 \alpha} = 2 g_A^e g_V^\alpha + \delta C_{1 \alpha}$ and $C_{2 \alpha} =
2 g_V^e g_A^\alpha + \delta C_{2 \alpha}$, where the standard model coupling
constants are
$g_{V,A}^f = Q^L_{w f} \pm Q^R_{w f}$ in terms of the left and right ($\alpha
=L,R$) weak charges
\begin{align}
\label{WeakCharge}
Q^\alpha_{w,f} = T_3 (f_\alpha) - Q(f) \sin^2 \theta_W \, .
\end{align}
The  ($\delta C_{i \alpha}$)
are inaccessible in other measurements, which gives PV-DIS
its unique quality.

The projected sensitivity of the SoLID experiment for an asymmetry discussed below
is $\delta A/A=\pm 0.005 ({\rm stat.})$ at an average $Q^2$ of 3.3 GeV$^2$
and an average $x$ of $\langle x \rangle =0.34$,
which sets the scale for
the size of acceptable theoretical uncertainties.
At this level of precision
several sources of systematic uncertainties can hamper a precise determination
of the $C_{i\alpha}$,
as discussed recently in Refs.~\cite{Hobbs:2008mm,Mantry:2010ki}. Some of the
most relevant are
uncertainties in leading-twist
parton distributions functions, in particular charge-symmetry violation (CSV),
contributions
from higher-twist correlation functions, and kinematical target-mass
corrections.
Far from being a nuisance higher-twist correlations encode very interesting and
yet
little known information on hadron structure. Therefore, all cases in which
leading-twist contributions are absent or reduced, such that one has a good
chance to
determine higher-twist ones are of great interest. If the relevant higher-twist
contributions are measurable with a given experimental sensitivity, as we will
claim
they are not in this case, one is in
a win-win situation: PV-DIS can be regarded either as a tool to find New
Physics,
in case the effects of the latter are prominent, or it can be seen
as a venue to access unknown aspects of strong interaction physics.

Parity violating weak interactions give rise to an asymmetry in the inclusive
cross
sections for scattering of left- and right-handed electrons off a deuteron
\begin{align}\label{asymmetry}
A_{RL}=\frac{d\sigma^R-d\sigma^L}{d\sigma^R+d\sigma^L}\,.
\end{align}
This is the main medium-energy observable in PV-DIS which will be scrutinized at
Jefferson
Lab \cite{Souder:2008zz,JLAB1,JLAB2}. Among all uncertainties of the theoretical
prediction of this asymmetry we will focus on the power suppressed
contributions. Two
recent studies of it reached somewhat different
conclusions~\cite{Hobbs:2008mm,Mantry:2010ki}. Our results turn out to be very
similar to
those from \cite{Mantry:2010ki}.

It was demonstrated by Bjorken and Wolfenstein
\cite{Bjorken:1978ry,Wolfenstein:1978rr}
that twist-four corrections to the asymmetry are due to a single (nonlocal)
four-quark
operator. The first estimates of the matrix element of the spin-two part of this
operator
  were obtained in the framework of the MIT bag model
\cite{Fajfer:1984um,Castorina:1985uw}. This technique was extended in
Ref.~\cite{Mantry:2010ki} to include the effects of higher spin operators. It
was found
that their effect is negligible within the model used. Renormalon analysis
offers yet
another technique to model the momentum fraction dependence of certain higher
twist matrix
elements \cite{DasWeb96,Ben99}. These renormalon-based studies
demonstrate~\cite{DasWeb96}
that higher-twist correlation functions (involving two quarks and a gluon) tend
to grow at
large $x$, i.e., like $(1 - x)^{-1}$, in qualitative agreement with experimental
measurements of electroweak structure functions~\cite{Kataev:1997nc}. However,
the
four-quark operators we consider are free from ultraviolet renormalons
\cite{GarKorRosTaf02} and thus this approach is not applicable. The absence of
renormalon
contributions might explain the qualitatively different behavior of such
correlators and
gluonic ones. In this work we calculate twist-four corrections employing a model
for the
nucleon wave functions in the light-cone formalism which was proposed by
\cite{Bolz:1996sw,Diehl:1998kh,BLMP}.

The paper is organized as follows: Sect.~\ref{sect:preliminaries} contains basic
definitions and notations.
In Sect.\ \ref{sect:twist4} we give a detailed discussion of power corrections
to the asymmetry (\ref{asymmetry}).
In Sect.\ \ref{sect:light-cone} the necessary ingredients of the light-cone
formalism are given. Results of our
calculation and our prediction for the twist-four corrections to the asymmetry
are collected in Sect.\ \ref{sect:results}.
Finally we give our conclusions. Several Appendices contain technical details
and
formulae left out in the body of the paper.

\section{Preliminaries}\label{sect:preliminaries}

\begin{figure}
\psfrag{k}[cc][cc][1.2]{$k$}
\psfrag{z}[cc][cc][1.2]{$k'$}
\psfrag{q}[cc][cc][1.2]{$q$}
\psfrag{p}[cc][cc][1.2]{$p$}
\psfrag{e}[cc][cc][1.2]{$e$}
\psfrag{ep}[cc][cc][1.2]{$e'$}
\psfrag{X}[cc][cc][1.2]{$X$}
\psfrag{D}[cc][cc][1.2]{$D$}
\psfrag{kp}[cc][cc][1.2]{$k'$}
\psfrag{ps}[cc][rc][1.2]{$p-p_s$}
\psfrag{Pps}[cc][rc][1.2]{$p_s$}
\psfrag{N}[cc][cc][1.2]{$N$}
\includegraphics[width=8cm,clip=true]{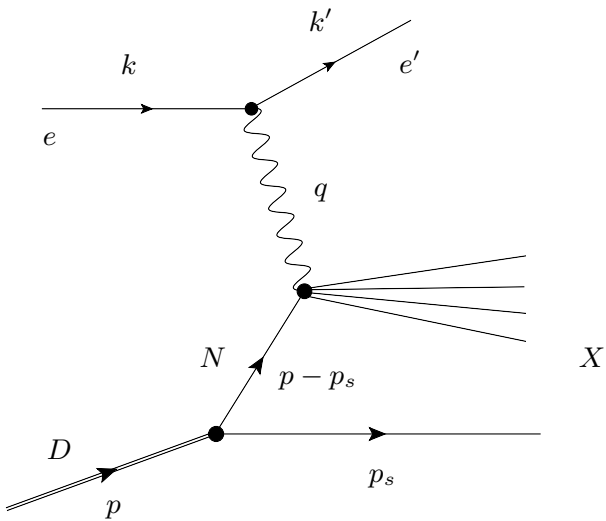}
\caption{\label{Kinematics} Kinematics in deep inelastic deuteron electron
scattering.}
\end{figure}

Let us briefly discuss the physical observables we will be analyzing below. The
cross section for polarized
electron scattering off an unpolarized deuteron target, with kinematics shown in
Fig.\ \ref{Kinematics}, is
given by the sum of three terms
\begin{align}\label{SigmaSum}
d\sigma^{L/R}=d\sigma^{L/R}_{ee}+d\sigma^{L/R}_{ww}+2d\sigma^{L/R}_{ew}
\end{align}
which describe the contributions due  to the electromagnetic and weak
interactions and their interference. Each term is a function of the standard
kinematical variables
\begin{align}
Q^2=-q^2,&&\nu=(p \cdot q),&& x=\frac{Q^2}{2\nu},&&y=\frac{(p \cdot q)}{(p \cdot
k)} ,
\end{align}
Each term in Eq.\ (\ref{SigmaSum}) is given by the convolution of a leptonic and
hadronic tensor. This reads in the  laboratory frame
\begin{align}\label{3sigma}
\frac{d\sigma_{ab}^{L/R}}{d\Omega dk'_0}=\frac{k'_0}{k_0}\,A_{ab}(Q^2) \,
(L^{L/R}_{ab})_{\mu\nu} W^{\mu\nu}_{ab}\,,
\end{align}
where the repeated Latin indices imply summation over electromagnetic and weak
exchanges $a,b=(e,w)$.
The coefficients
\begin{align*}
A_{ee}(Q^2)&=\frac{2\alpha^2}{Q^4}\,, &&
A_{ew}(Q^2)=\frac{\sqrt{2}G_F\alpha}{\pi Q^2}\,,\\
A_{ww}(Q^2)&=\frac{G_F^2}{\pi^2}
\end{align*}
encode the products of gauge boson propagators and interaction
strengths.
The leptonic tensor admits the conventional form
\begin{align}\label{}
(L^{L/R}_{ab})_{\mu\nu}=Q^{L/R}_a\,Q^{L/R}_b \ell_{\mu\nu}^{L/R}\, ,
\end{align}
which is a product of the electromagnetic (weak) charge \footnote{For brevity,
we omit the subscript from $Q$
labeling the particle species.} $Q_{e(w)}^{L/R}$ for the left (right) handed
electron
\begin{align*}\label{}
Q_e^{L/R}=-1\,,&& Q_{w}^{L}=-\frac12+\sin^2\theta_W\,,
&&Q_{w}^{R}=\sin^2\theta_W\, ,
\end{align*}
and
\begin{align}\label{ell}
\ell^{L/R}_{\mu\nu}=  k_\mu k'_\nu+k'_\mu k_\nu-g_{\mu\nu}(k \cdot k')
\pm
i\varepsilon_{\mu\nu\rho\sigma} k^\rho {k'}^\sigma\,.
\end{align}
The hadronic tensor $W^{\mu\nu}_{ab}$ is the deuteron matrix element of the
product of currents
\begin{multline}\label{}
W^{\mu\nu}_{ab}(p,q)=\frac{1}{8\pi M_D}
\int d^4z \,e^{iq \cdot z}\\
\times\langle{D(p)|\Big\{j^\mu_a(z)j^\nu_b(0)+j^\mu_b(z)j^\nu_a(0)\Big\}|D(p)}\rangle
\,,
\end{multline}
where $M_D$ is the deuteron mass and
averaging over deuteron polarizations is implied. The electromagnetic and
neutral quark current
are defined as (cf.\ Eq.\ (\ref{WeakCharge}))
\begin{align}\label{}
j^\mu_e=\bar q Q\gamma^\mu q\,,&&
j^\mu_w=\bar q_L\tau_3 \gamma^\mu q_L-\sin^2\theta_W j^{\mu}_e\,,
\end{align}
where $q=(u,d)$. It was demonstrated by Bjorken~\cite{Bjorken:1978ry} that if
one
assumes valence quark dominance in the region $x>0.4$ and neglects all sea quark
and
isospin breaking effects (which should be justified for large virtual mass
$Q^2$),
the asymmetry~(\ref{asymmetry})
becomes free of hadronic physics contaminations and is given
by the Cahn-Gilman formula~\cite{Cahn:1977uu}
\begin{multline}\label{CG}
A_{RL}=-\frac{G_F Q^2}{2\sqrt{2}\pi\alpha}\frac{9}{10}
\biggl[
\left(1-\frac{20}{9}\sin^2\theta_W\right)
\\
+(1-4\sin^2\theta_W)\frac{1-(1-y)^2}{1+(1-y)^2}
\biggr]\,.
\end{multline}
New Physics is best parameterized by allowing for non-standard values for the
coefficients
$C_{i\alpha}$, which are reintroduced in Eq.(\ref{CG}) by replacing
\begin{align*}
1-\ft{20}{9}\sin^2\theta_W
&
\to - \ft23 (2 C_{1u}- C_{1d})
\, , \\
1-4\sin^2\theta_W
&
\to - \ft23 (2 C_{2 u} - C_{2 d})
\, .
\end{align*}

However, the assumptions leading to vanishing hadronic effects are only valid
approximately
and have to be abandoned in the analysis of high precision experiments.
The main hadronic effects are caused by CSV and power suppressed correlators.
The central point behind our work, and that of others, is that these effects
have a strong $x$
dependence which allows, if precisely known, to isolate and subtract them
and
thus to increase the sensitivity of experiments like SoLID to New Physics.
Thus one has to go beyond leading approximations and has to take into account
higher-order electro-weak effects, sea quark effects,
target mass and higher-twist corrections at least at a level matching the
accuracy of
experimental measurements.

CSV arises from isospin violation of $u$ and $d$ quark distributions in the
proton and neutron,
i.e., by $\delta u = u_p- d_n\neq 0 $ and $\delta d = d_p - u_n\neq 0$. Modern
global analysis
of parton distribution functions incorporate CSV effects, which are found to
become more
significant as $x$ decreases~\cite{Martin:2003sk},
$R_{\rm\scriptscriptstyle CSV} \sim (\delta u - \delta d)/(u + d) \sim (1 -
x)^4 \sqrt{x}$.
CSV effects might explain a significant fraction of
the discrepancy between the NuTeV results~\cite{NuTeV03} and predictions based
on the
standard model and isospin symmetry.

The other source of corrections are power suppressed contributions from multi-
particle correlation
functions. Obviously the nucleon wave function is a complex state containing
many highly entangled
Fock states, only partially characterized by parton distribution functions.
The isolation and determination of specific multiple-field correlators is the
logical next step to
explore hadrons and is therefore of great interest in its own right.
In contrast to mere one particle probability distributions, they contain
information on
relative phases. As they are typically power suppressed, high luminosity
experiments at medium
large $Q^2$ are needed to extract them. These are requirements which are
perfectly fit
by Jefferson Lab, especially after the energy upgrade.

\section{Twist four corrections}\label{sect:twist4}
In the region of  low transferred momentum $Q^2\ll M_W^2$ one has
$d\sigma_{ww}\ll d\sigma_{ew}\ll d\sigma_{ee}$
and the asymmetry takes the form
\begin{align}\label{}
A_{RL}=\frac{d\sigma_{ew}^R-d\sigma_{ew}^L}{d\sigma_{ee}}\,,
\end{align}
where we took into account that $d\sigma^L_{ee}=d\sigma^R_{ee}\equiv
d\sigma_{ee}$.

Introducing the scalar, isovector and axial isovector currents
\begin{align}
S^\mu=\frac12\bar q\gamma^\mu  q\,, && V^\mu=\bar q\gamma^\mu \tau^3 q\,, &&
A^\mu=\bar q\gamma^\mu \gamma_5\tau^3 q\,
\end{align}
one can represent the  electromagnetic (weak) hadronic tensors as follows
\begin{align}\label{}
W^{\mu\nu}_{ee}(p,q)=&W^{\mu\nu}_{V}(p,q)+\frac19 W^{\mu\nu}_{S}(p,q)\,,\notag\\
W^{\mu\nu}_{ew}(p,q)=&\left(\frac12-
\sin^2\theta\right)W^{\mu\nu}_{V}(p,q)\notag\\
&-
\frac19\sin^2\theta \,W^{\mu\nu}_{S}(p,q)-\frac12 W^{\mu\nu}_A(p,q)\,,
\end{align}
where
\begin{align}\label{WWW}
W^{\mu\nu}_{V}(p,q)=&\frac{1}{4\pi M_D}
\int d^4z e^{iq \cdot z}
\langle{D(p)|V^\mu(z)V^\nu(0)|D(p)}\rangle\,,
\notag\\[2mm]
W^{\mu\nu}_{S}(p,q)=&\frac{1}{4\pi M_D}
\int d^4z e^{iq \cdot z}
\langle{D(p)|S^\mu(z)S^\nu(0)|D(p)}\rangle\,,\notag\\
W^{\mu\nu}_{A}(p,q)=&\frac{1}{8\pi M_D}\int d^4z e^{iq \cdot z}
\langle D(p)|A^\mu(z)V^\nu(0)
\notag\\
&\hspace*{2cm}+V^\mu(z)A^\nu(0) |D(p)\rangle\,.
\end{align}
Here we  took  into account that the deuteron matrix elements of nonsinglet
terms, i.e., involving
the product of isovector and isosinglet currents $V S$ and $A S$, vanish by
isospin symmetry since
the deuteron is an isoscalar state. Keeping only twist-two terms in the OPE
expansion of the
hadronic tensors~(\ref{WWW}) one arrives at the Cahn-Gilman formula~(\ref{CG}),
the first and
the second term in the square brackets in~(\ref{CG}) arise from vector-vector
($W^{v}_{ew}$) and
axial-vector  ($W^{a}_{ew}$) correlators, respectively. The corrections to the
Cahn-Gilman
formula can be parameterized as follows
\begin{align}\label{}
A_{RL}=-\frac{G_F Q^2}{2\sqrt{2}\pi\alpha}\frac{3}{5}
\biggl[
\tilde a_1+\tilde a_2\,\frac{1-(1-y)^2}{1+(1-y)^2}
\biggr]\,,
\end{align}
where ($i =1,2$)
\begin{align}\label{}
\tilde a_i &= - (2 C_{i u}- C_{i d}) \left[1+R_i \right]\,.
\end{align}
Here the  functions $R_i$ ($i = 1,2$) alluded to before receive contributions
from  several
sources of hadronic effects. The precision measurement of the mixing angle at
low $Q^2$ gives  $\sin^2\theta_W
\simeq 0.2397$ \cite{Anthony:2005pm}.  Thus  the axial current contribution
($\tilde a_2$) to the asymmetry
is relatively small and we will focus on the calculation of twist-four
corrections to $\tilde a_1$.  They can
be easily identified. Indeed, neglecting effects  of isospin breaking one gets
(see Ref.~\cite{Bjorken:1978ry})
\begin{multline}\label{}
\langle{D|S^\mu(z)S^\nu(0)-V^\mu(z)V^\nu(0)|D}\rangle=
\\
=\frac12\langle{D|\bar u(z)\gamma^\mu u(z)\,\bar d(0)\gamma^\nu
d(0)+(u\leftrightarrow d)|D}\rangle\,.
\end{multline}
The expansion of the operator at the right-hand side of  this equation starts
from twist-four.
In terms of
\begin{align}\label{}
& W^{\mu\nu}_{ud}(p,q) =\frac{1}{8\pi M_D}
\int d^4z e^{iq \cdot z}
\\
& \quad\times \langle{D(p)|\bar u(z)\gamma^\mu u(z)\,\bar d(0)\gamma^\nu
d(0)+(u\leftrightarrow d)
|D(p)}\rangle \nonumber
\end{align}
we define the structure functions $F_{i=1,2}^a$ as coefficients in the tensor
decomposition
\begin{align}\label{}
M_D W^{\mu\nu}_{a}(p,q)=&\left(-g^{\mu\nu}+\frac{q^\mu q^\nu}{q^2}\right)F_1^a
\\
&+\frac1{\nu}
\left(p^\mu-\frac{(pq)}{q^2} q^\mu\right)\left(p^\nu-\frac{(pq)}{q^2}
q^\nu\right) F_2^{a}\,,\notag
\end{align}
where the index runs over $a=V,S, ud$. The twist-four contribution to  $R_1$
takes the form
\begin{align}\label{}
R_1^{\rm tw-4}=-\frac1{10(1-
\frac{20}9\sin^2\theta_W)}\frac{\mathcal{F}^{ud}}{\mathcal{F}^{S}}\,,
\end{align}
where
\begin{align}\label{}
\mathcal{F}^a=xy F_1^a-\left[1-\frac1{y}+\frac{x M_D}{2E}\right ]F_2^a\,.
\end{align}
Keeping in $\mathcal{F}^S$ and $\mathcal{F}^{ud}$ the dominant contributions
only, i.e,
twist-two and twist-four, respectively, and taking into account that they both
satisfy  the
Callan-Gross relation $F_2=2x F_1$, one finds
\begin{align}\label{}
\frac{\mathcal{F}^{ud}}{\mathcal{F}^{S}}\simeq\frac{{F}_1^{ud}}{{F}_1^{S}}\,.
\end{align}
The expression for $F_1^S$ at lowest order of perturbation theory is given by
the sum of
parton densities in the deuteron
\begin{align}\label{}
F_1^S(x)=\frac18\Big[u_D(x)+d_D(x)+\bar u_D(x)+\bar d_D(x)\Big]\,,
\end{align}
where  as usual $\bar q_D(x)=-q_D(-x)$. The quark distribution functions are
defined by the matrix
elements of nonlocal light-cone operators,
\begin{align}\label{}
\langle D|\bar q(z)\slashed{z}q(-z)|D\rangle=2(p \cdot z)\int_{-1}^1dx e^{2i (p
\cdot z)x} q_D(x)\,.
\end{align}
To evaluate $F_1^{ud}$ we represent the hadronic tensor $W^{\mu\nu}_{ud}$ via
the dispersion
relation as a time-ordered product of electroweak currents
\begin{align}\label{Wud}
&W^{\mu\nu}_{ud}(p,q)=\text{Im}\biggl[\frac{i}{4\pi M_D}
\int d^4z\, e^{iq \cdot z}
\\
&\quad \times\langle{D(p)|T\{\bar u(z)\gamma^\mu u(z)\,\bar d(0)\gamma^\nu
d(0)+(u\leftrightarrow d)\}|D(p)}\rangle\biggr]\,
\notag
\end{align}
and make use of the operator product expansion~\cite{Balitsky:1987bk}
\begin{align}\label{Texp}
T\Big\{\bar u(z)\gamma_\mu u(z) &\,\bar d(-z)\gamma_\nu d(-z)+(u\leftrightarrow
d)\Big\}^{\rm tw-4}
\notag\\
=&\frac{\alpha_s}{16\pi i}\biggl\{
-\log z^2 \partial_\mu\partial_\nu \int_0^1 du \frac{\bar u}{u^2}
\mathcal{Q}(uz)
\notag\\
&
+\frac1{z^2}
S_{\mu\alpha\nu\beta} z^\alpha\partial^\beta\int_0^1 \frac{du}{u}
\mathcal{Q}(uz)
\biggr\}\, ,
\end{align}
where $S_{\mu\alpha\nu\beta} = g_{\mu\alpha}  g_{\nu\beta} + g_{\nu\alpha}
g_{\mu\beta} - g_{\mu\nu}
g_{\alpha\beta} $.

The operator $\mathcal{Q}$ ($\mathcal{Q}_2$ in the notations of
Ref.~\cite{Balitsky:1987bk})
is given by the following expression
\begin{align}\label{Q2}
\mathcal{Q}(z)=&
i\int_{-1}^1 dv \int_{-1}^v dt\Biggl[\Pi_{12}^-\Pi_{34}^-\mathcal{Q}_V(1,v,t,-1)
\,
\notag\\
&+\Pi_{12}^+\Pi_{34}^+\,\mathcal{Q}_A(1,v,t,-1)
\Biggr]\,
+(z\leftrightarrow -z)\,.
\end{align}
Here
\begin{align}\label{QVA}
\mathcal{Q}_A(a)=&
\Big(\bar u(a_1z) t^a \slashed{z}\gamma_5 u(a_2z)\Big)\,
\Big(\bar d(a_3z) t^a \slashed{z}\gamma_5d(a_4z)\Big)\,,
\notag\\
\mathcal{Q}_V(a)=&
\Big(\bar u(a_1z) t^a \slashed{z} u(a_2z)\Big)\,\Big(\bar d(a_3z) t^a
\slashed{z}d(a_4z)\Big)\,,
\end{align}
and $\Pi^{\pm}_{ik}=(1\pm P_{ik})$, where
$P_{ik}$ is  the permutation operator, e.g.,
$P_{12}\mathcal{Q}_V(a_1,a_2,a_3,a_4)=\mathcal{Q}_V(a_2,a_1,a_3,a_4)$.
For later convenience we rewrite~(\ref{Q2}) as follows
\begin{align}\label{Q22}
\mathcal{Q}(z)=&
i\int_{-1}^1 dv \int_{-1}^v dt\Big[\widehat{\mathcal{Q}}_+(1,v,t,-1)-
\widehat{\mathcal{Q}}_-(1,v,t,-1)]\,,
\end{align}
where
\begin{align}\label{}
\widehat{\mathcal{Q}}_+(a)=&(1+P_{12}P_{34})(1+P_{14}P_{23})
\mathcal{Q}_+(a)\,,\notag\\
\widehat{\mathcal{Q}}_-(a)=&(P_{12}+P_{34})(1+P_{14}P_{23}) \mathcal{Q}_-(a)
\end{align}
and
\begin{align}\label{Qpm}
\mathcal{Q}_\pm(a)=\mathcal{Q}_V(a)\pm \mathcal{Q}_A(a)\,.
\end{align}
Let us define the twist-four distribution $\widetilde{\mathcal{Q}}_D(x)$  as a
deuteron matrix element of the
operator $\mathcal{Q}$
\begin{align}\label{}
\langle{D|\mathcal{Q}(z)|D\rangle}=i\int_{-1}^1dx \,e^{2i (p\cdot
z)x}\,\widetilde{\mathcal{Q}}_D(x)\,.
\end{align}
It follows from (\ref{Q2}) and (\ref{Qpm}) that $\widetilde{\mathcal{Q}}_D(x)$
is an even function
of $x$ with vanishing first moment,
$$
   \int_{-1}^1dx\,\widetilde{\mathcal{Q}}_D(x)=0.
$$
Inserting~(\ref{Texp}) and(\ref{Q2}) into (\ref{Wud}) one finds after some
algebra
\begin{align}\label{}
F_1^{ud}(x)=-\frac{\alpha_s\pi}{4Q^2}\, x\, \widetilde{\mathcal{Q}}_D(x)\,.
\end{align}
Then, keeping in $F_1^S$ the valence quark contribution only  we obtain the
following expression
for  the twist-four  correction to the asymmetry
\begin{align}\label{}
R_1^{\rm tw-4}=\frac1{Q^2}\frac{\alpha_s\pi}{5(1-
\frac{20}9\sin^2\theta_W)}\frac{x\,
\widetilde{\mathcal{Q}}_D(x)}{u_D(x)+d_D(x)}\,.
\end{align}

The deuteron is a weakly coupled state of the proton and neutron with the
binding energy
$E_B\simeq 2.2\,\text{MeV}$. In the incoherent impulse approximation its
hadronic tensor
in the deuteron's rest frame can be represented as ~\cite{Atwood:1972zp}%
\begin{align}\label{DN}
W_{\mu\nu}^D(p,q)\simeq&\int \frac{d^3 \bit{p}_s}{(2\pi)^3 E_{p_s}/M_N}|f(
\bit{p}_s)|^2\notag\\
&\times
\Big(W_{\mu\nu}^{(p)}(p - p_s,q)+W_{\mu\nu}^{(n)}(p - p_s,q)\Big)\,,
\end{align}
where $p_s=(E_{p_s},\bit{p}_s)$ and the integration is performed over the
spectator three-momentum
$\bit{p}_s$, see Fig.\ \ref{Kinematics}. Here $f( \bit{p}_s )$ is the deuteron
wave function in its rest frame,
normalized as $\Big[(2\pi)^{-3}\int{d^3 \bit{p}_s}\, |f(\bit{p}_s)|^2=1\Big]$
and $W_{\mu\nu}^{(p(n))}$ are
the proton (neutron) hadronic tensors. The function $f(\bit{p}_s)$ is strongly
peaked at $\bit{p}_s=0$
\cite{Atwood:1972zp}. Thus one can simplify the above expression by neglecting
terms of order $\sim |
\bit{p}_s|/M_N$ and higher under the integral. Then one finds
\begin{align}\label{WDN}
W_{\mu\nu}^D(p,q)\simeq W_{\mu\nu}^{(p)}(p/2,q)+W_{\mu\nu}^{(n)}(p/2,q)\,,
\end{align}
and as a consequence $d\sigma_d\simeq d\sigma_p+d\sigma_n$. Then Eq.\
(\ref{WDN}) yields the
 following relation between the structure functions of deuteron and nucleons,
$$
F_2^d(x/2)\simeq F_2^p(x)+F_2^n(x).
$$
It turns out that this approximation overestimates the deuteron structure
function by $5\div 10\%$
\cite{Atwood:1972zp,Schmidt:1977hs}. This is acceptable for our purposes. For
the parton densities
the corresponding relation reads (cf. Eqs.~(47) and (48) in
Ref.~\cite{Mantry:2010ki})
\begin{align}\label{}
\frac12 q_D(x/2)\simeq&q_p(x)+q_n(x)\,.
\end{align}
Similarly, defining the proton (neutron) twist-four distributions  by
\begin{align}\label{}
\langle{N|\mathcal{Q}(z)|N\rangle}=i\int_{-1}^1dx \,e^{2i (p \cdot z)x}\,
\widetilde{\mathcal{Q}}_p(x)\,
\end{align}
one gets for the deuteron twist-four function $\widetilde{ \mathcal{Q}}_{D}(x)$
\begin{align}
\frac14\widetilde{ \mathcal{Q}}_{D}({x}/{2})=&
\widetilde{ \mathcal{Q}}_{p}(x)+\widetilde{ \mathcal{Q}}_{n}(x)=
2\widetilde{ \mathcal{Q}}_{p}(x)\,.
\end{align}
Here we took into account that
$\widetilde{ \mathcal{Q}}_p(x)=\widetilde{ \mathcal{Q}}_n(x)$ due to isospin
symmetry.

We also define the nucleon twist-four distribution ${\mathcal{Q}}_{\pm}(\xi)$
(and
similarly $\widehat{\mathcal{Q}}_\pm(\xi)$)
by
\begin{align}\label{QVAD}
\langle{N|\mathcal{Q}_{\pm}(a)|N}\rangle=&(p \cdot z)^2\int \mathcal{D}\xi e^{-
i(p \cdot z)\sum_k a_k\xi_k}
{\mathcal{Q}}_{\pm}(\xi)\,,
\end{align}
where $\xi$ cumulatively denotes the array of four variables $\xi = (\xi_1,
\xi_2, \xi_3, \xi_4)$ and
the integration measure stands for $\mathcal{D}\xi=\prod_{k=1}^4d\xi_k\,
\delta(\sum_i\xi_i)$. Then
it follows from Eq.~(\ref{Q22}) that
\begin{multline}\label{Q2I}
\widetilde{\mathcal{Q}}_p(x)=\int \frac{\mathcal{D}\xi}{\xi_2\xi_3(\xi_2+\xi_3)}
\Big\{(\xi_2+\xi_3)\delta(x+\xi_1+\xi_2)
\\
-\xi_3\delta(x+\xi_1)-\xi_2\delta(\xi_4-x)
\Big\}[\widehat{\mathcal{Q}}_+(\xi)-\widehat{\mathcal{Q}}_-(\xi)]\,.
\end{multline}
%

\section{Nucleon light-cone wave functions}\label{sect:light-cone}

Our lack of information on
the magnitude of higher-twist matrix elements is the main obstacle for a
quantitative analysis of power-suppressed contributions to hadronic cross
sections.
Hadron structure models provide estimates for the size of nonperturbative matrix
elements, but
their predictions vary strongly. This is understandable in view of the fact that
confinement
is incorporated rather differently.  The first estimates of twist-four
corrections to the
asymmetry (\ref{asymmetry}) were obtained within the MIT bag
model~\cite{Fajfer:1984um,Castorina:1985uw} which incorporates confinement quite
ad hoc, (see also
Refs.~\cite{Mantry:2010ki,Sacco:2009kj} for recent developments). In this work
we use another
approach, the light-cone formalism~\cite{Lepage:1980fj},
for the evaluation of twist-four corrections.

In the light-cone formalism the nucleon is represented by  a superposition of
multi-parton Fock state wave
functions. The latter are functions of the parton longitudinal momentum
fractions $x_i$, transverse momenta
$\bit{k}_{\perp i}$, and parton helicities. The light-cone wave functions
(LCWFs)  are eigenfunctions of the
QCD Hamiltonian quantized in the light-cone
gauge~\cite{Kogut:1969xa,Brodsky:1997de}.
Models for LCWFs
of various degree of sophistication have been  considered in different context
in the vast literature on the
subject, see, e.g.,
Refs.~\cite{Lepage:1980fj,Bolz:1994hb,Bolz:1996sw,Diehl:1998kh,Ji:2003yj,Pasquini:2008ax}.
In this work we will follow the formalism developed in
Refs.~\cite{Bolz:1994hb,Bolz:1996sw,Diehl:1998kh,BLMP}
and will take into account only the lowest components of the nucleon LCWFs: the
three quark and three-quark-gluon
component. The details of the light-cone formalism relevant for our further
discussion are collected in
Appendix~\ref{app:light-cone}.

The three quark component of the nucleon state is parameterized in terms of
corresponding LCWF
$ \Psi_{123}^{(0)}$ as follows
\begin{multline}\label{Ansatz}
\ket{p,+}_{3q}
=-\frac{\epsilon^{ijk}}{\sqrt{6}}\int[\mathcal{D}X]_3
\Psi_{123}^{(0)}(X)\times\,
\\
\Bigl(u_{i\uparrow}^\dagger(1)
u_{j\downarrow}^\dagger(2)d_{k\uparrow}^\dagger(3)-
u_{i\uparrow}^\dagger(1)d_{j\downarrow}^\dagger(2)u_{k\uparrow}^\dagger(3)
\Bigr)\ket{0}\,.
\end{multline}
Here and below for notational simplicity arguments like $\ell$ in
$u^\dagger_{i\uparrow}(\ell)$, stand for the
collection of all relevant arguments, i.e., $u^\dagger_{\uparrow i}(\ell)
=u^\dagger_{\uparrow i}(x_\ell, \bit{k}_{\perp\ell})$.
The creation (annihilation) operators  of a quark
with helicity $\lambda$ and momentum $p$ satisfy the commutation
relation~(\ref{bbdd}).
As usual, the momentum fraction $x_i$ is defined as ratio of the longitudinal
(i.e., ``+") momentum
of the $i-$th parton and the one of the nucleon. The integration measure has the
following form
\begin{align}\label{measurek}
[\mathcal{D}X]_N=&\frac{1}{\sqrt{x_1\ldots x_N}}[dx]_N[d^2 \bit{k}_\perp]_N\,,
\notag
\\
[dx]_N=&\prod_{i=1}^N dx_i\,\delta(1-\sum x_i)\,,\notag
\\
[d^2 \bit{k}_\perp]_N=&\frac1{(16\pi^3)^{N-1}}\prod_{i=1}^N d^{2} \bit{k}_{\perp
i}\,
\delta^{(2)}\left(\sum \bit{k}_{\perp i}\right)\,.
\end{align}

\vspace{3mm}

Here we accept the Bolz-Kroll  ansatz~\cite{Bolz:1996sw} for the function
$\Psi_{123}^{(0)}$
\begin{align}\label{Psi0}
\Psi_{123}^{(0)}=&\frac{f_N}{4\sqrt{6}}\,\phi(x_1,x_2,x_3)\, \Omega_3(a_3,x_i,
\bit{k}_{\perp i})\,.
\end{align}
The transverse momentum dependence is encoded in the function $\Omega_N$
\begin{align}
\Omega_N(a_N,x_i, \bit{k}_{\perp i})=\frac{(16\pi^2 a_N^2)^{N-1}}{x_1x_2\ldots
x_N}
\exp\left[-a_N^2\sum_i \bit{k}_{\perp i}^2/x_i  \right]\,
\label{eq:Omega}
\end{align}
which is normalized such that
\begin{eqnarray}
&&\int [d^{2} \bit{k}_{\perp}]_N\Omega_N(a_N,x_i, \bit{k}_{\perp i})\,=\,1\,,
\nonumber\\
&&\int [d^{2} \bit{k}_{\perp}]_N\Omega^2_N(a_N,x_i, \bit{k}_{\perp
i})\,=\,\frac{\rho_N}{x_1\ldots x_N}\,,
\label{eq:Omeganorm}
\end{eqnarray}
where $ \rho_N = (8\pi^2 a_N^2)^{N-1}$. The function $\phi(x_i)$, entering
(\ref{Psi0}), depends only
on the longitudinal momentum fractions of constituent partons and is related to
the leading-twist, i.e.,
twist-three, nucleon distribution amplitude, namely,
\begin{align}\label{phi3}
\phi(x_1,x_2,x_3) =\Phi_3(x_1,x_2,x_3;\mu_0)\,.
\end{align}
Here $\Phi_3(x)$ is the twist-three nucleon distribution amplitude defined at
the low-energy scale
$\mu_0=1\text{GeV}$. We use the following ansatz for $\Phi_3(x)$
\cite{Bolz:1996sw}

\begin{align}\label{Phi3}
\Phi_3(x)=60\, x_1x_2x_3\,(1+3x_1)\,,
\end{align}
which emerges from the truncation of the conformal partial wave expansion after
the lowest few terms.
The normalization constant $f_N$ in Eq.\ (\ref{Psi0}) is determined by the
matrix element of the
corresponding local three-quark operator. The analysis within the framework of
QCD sum rules
\cite{SVZ79} yields in the following estimate for $f_N$
\cite{Chernyak:1984bm,King:1986wi,Chernyak:1987nt,Braun:2000kw,Gruber:2010bj}
at the scale $\mu_0=1\,\text{GeV}$
\begin{equation}
 f_N  = (5.0\pm 0.5)\times 10^{-3}\,\, \text{GeV}^2.
\label{eq:fN}
\end{equation}
On the other hand, the parameter $a_3$ determines the smearing of the wave
function in the transverse
plane and, e.g., the average quark transverse momentum. Following
Ref.~\cite{BLMP} we take
$a_3=0.73\, \text{GeV}^{-1}$ in our estimates. With this set of parameters, the
contribution of
the three-quark Fock state to the norm of the nucleon state is about 17\%,
\begin{align}\label{}
P_{3q}=\frac{435}{112}f_N^2\rho_3\simeq 0.17\,.
\end{align}
The four-parton quark-gluon contributions with zero angular momentum to the
nucleon states have the
following form~\cite{BLMP}
\begin{widetext}
  \begin{align}\label{gminus}
\ket{p,+}_{uudg_\downarrow}
=&\epsilon^{ijk}\int[\mathcal{D}X]_4
\,\Psi^\downarrow_{1234}(X)\,
a_{\downarrow}^{a,\dagger}(4)\,
[t^a u_{\uparrow}(1)]_i^\dagger\, u^\dagger_{j\uparrow}(2))
\,d_{k\uparrow}^\dagger(3)
\ket{0}\,,
\notag\\
\ket{p,+}
_{uudg^\uparrow}=&\epsilon^{ijk} \int[\mathcal{D}X]_4\Big\{
\Psi^{\uparrow(1)}_{1234}(X)\,
[t^au_{\downarrow}(1)]_i^\dagger
\Bigl(u_{j\uparrow}^\dagger(2)d_{k\downarrow}^\dagger(3)-
d_{j\uparrow}^\dagger(2)
u_{k\downarrow}^\dagger(3)
\Bigr)a_{\uparrow}^{a,\dagger}(4)
\notag\\
&{}\hspace*{10mm}
+\Psi^{\uparrow(2)}_{1234}(X)
u^\dagger_{i\downarrow}(1)\Big( [t^a u_{\downarrow}(2)]^\dagger_j\,
d^\dagger_{k\uparrow}(3)-
[t^a d_{\downarrow}(2)]^\dagger_j\, u^\dagger_{k\uparrow}(3)\Big)
a_{\uparrow}^{a,\dagger}(4)
\Biggl\} \ket{0}\,,
\end{align}
\end{widetext}
where the four-parton LCWFs are again taken in the Bolz-Kroll form
\begin{align}\label{PhiPsi}
\Psi^\downarrow_{1234}=& \frac1{\sqrt{2x_4}}  \phi_g(x_1,x_2,x_3,x_4)\,
\Omega_4(a_g,x_i, \bit{k}_{\perp i})\,,
\nonumber\\
\Psi^{\uparrow(1)}_{1234}=& \frac1{\sqrt{2x_4}}
\psi_g^{(1)}(x_1,x_2,x_3,x_4)\,
\Omega_4(a_g,x_i, \bit{k}_{\perp i})\,,
\nonumber\\
\Psi^{\uparrow(2)}_{1234}= &  \frac1{\sqrt{2x_4}}
\psi_g^{(2)}(x_1,x_2,x_3,x_4)\,
\Omega_4(a_g,x_i, \bit{k}_{\perp i})\,.
\end{align}
The functions $\phi_g,\psi_g^{(i)}$ which depend on
the light-cone momentum fractions of the partons can be expressed in terms of
the twist-four quark-gluon nucleon distribution amplitudes introduced in Ref.\
\cite{Braun:2008ia},
\begin{eqnarray}\label{psiPsi}
\lefteqn{{g}\,\phi_g(x_1,x_3,x_2,x_4)=}
\nonumber\\&=&
-\frac{m_N}{96}\Big[2\Xi_4^g(x_1,x_2,x_3,x_4) + \Xi_4^g(x_2,x_1,x_3,x_4)\Big]\,,
\notag\\
\lefteqn{{g}\,\psi_g^{(1)}(x_1,x_2,x_3,x_4)=}
\nonumber\\&=&
-\frac{m_N}{48}\Big[\Psi_4^g(x_2,x_1,x_3,x_4)+\frac12\Phi_4^g(x_1,x_2,x_3,x_4)
\Big]\,,
\notag\\
\lefteqn{{g}\,\psi_g^{(2)}(x_1,x_3,x_2,x_4)=}
\nonumber\\&=&
\frac{m_N}{48}
\Big[\Phi_4^g(x_1,x_2,x_3,x_4)+\frac12\Psi_4^g(x_2,x_1,x_3,x_4)\Big]\,.
\end{eqnarray}
Keeping only the lowest terms in the
conformal expansion of the corresponding distribution amplitudes one arrives at
the following
expressions~\cite{BLMP}
\begin{align}\label{WFG}
g\phi_g (x_1, x_2, x_3, x_4)=&-210 m_N\lambda_1^g \, x_1x_2x_3
x_4^2\,,\notag\\[2mm]
g\psi_g^{(1)} (x_1, x_2, x_3, x_4)=&-105 m_N (\lambda_2^g+\lambda_3^g)\,
x_1x_2x_3 x_4^2\,,
\notag\\[2mm]
g\psi_g^{(2)} (x_1, x_2, x_3, x_4)=&-105 m_N (\lambda_2^g-\lambda_3^g)\,
x_1x_2x_3 x_4^2\,.
\end{align}
The sum rule technique was found to give the following estimates for the
coupling
constants $\lambda_k^g$ at low energy scale $1\,\text{GeV}$~\cite{BLMP}
\begin{align}
    \lambda_1^g = &  (2.6\pm 1.2)\cdot 10^{-3}\text{GeV}^2\,,
\notag\\
    \lambda_2^g = &  (2.3\pm 0.7)\cdot 10^{-3}\,\text{GeV}^2\,,
\notag\\
    \lambda_3^g = &  (0.54\pm  0.21)\cdot 10^{-3}\,\text{GeV}^2\,.
\label{eq:lambdag}
\end{align}
We choose $a_g=a_3/2^{1/6}=0.65\, \text{GeV}^{-1}$ and $\alpha_s=0.5$ at the
scale $1$ GeV
which results in the following probabilities for the quark-gluon components
within the nucleon
state~\cite{BLMP}
\begin{align}
P_{g^\downarrow} =& \frac{35}{8g^2}m_N^2\rho_4 (\lambda^g_1)^2
\simeq 0.15\,,
\notag\\
P_{g^\uparrow}  =& \frac{105}{16g^2}m_N^2\rho_4
\Big[(\lambda^g_2)^2 + (\lambda^g_3)^2\Big]\,
\simeq 0.185\,.
\end{align}

\section{Results and Discussion}\label{sect:results}

\begin{figure}[htb]
\psfrag{x}[cc][cc][1.2]{$x$}
\psfrag{R}[cc][cc][1.2]{$a_3^2\, \widetilde{Q}_p(x)$}
\includegraphics[width=0.44\textwidth,clip=true]{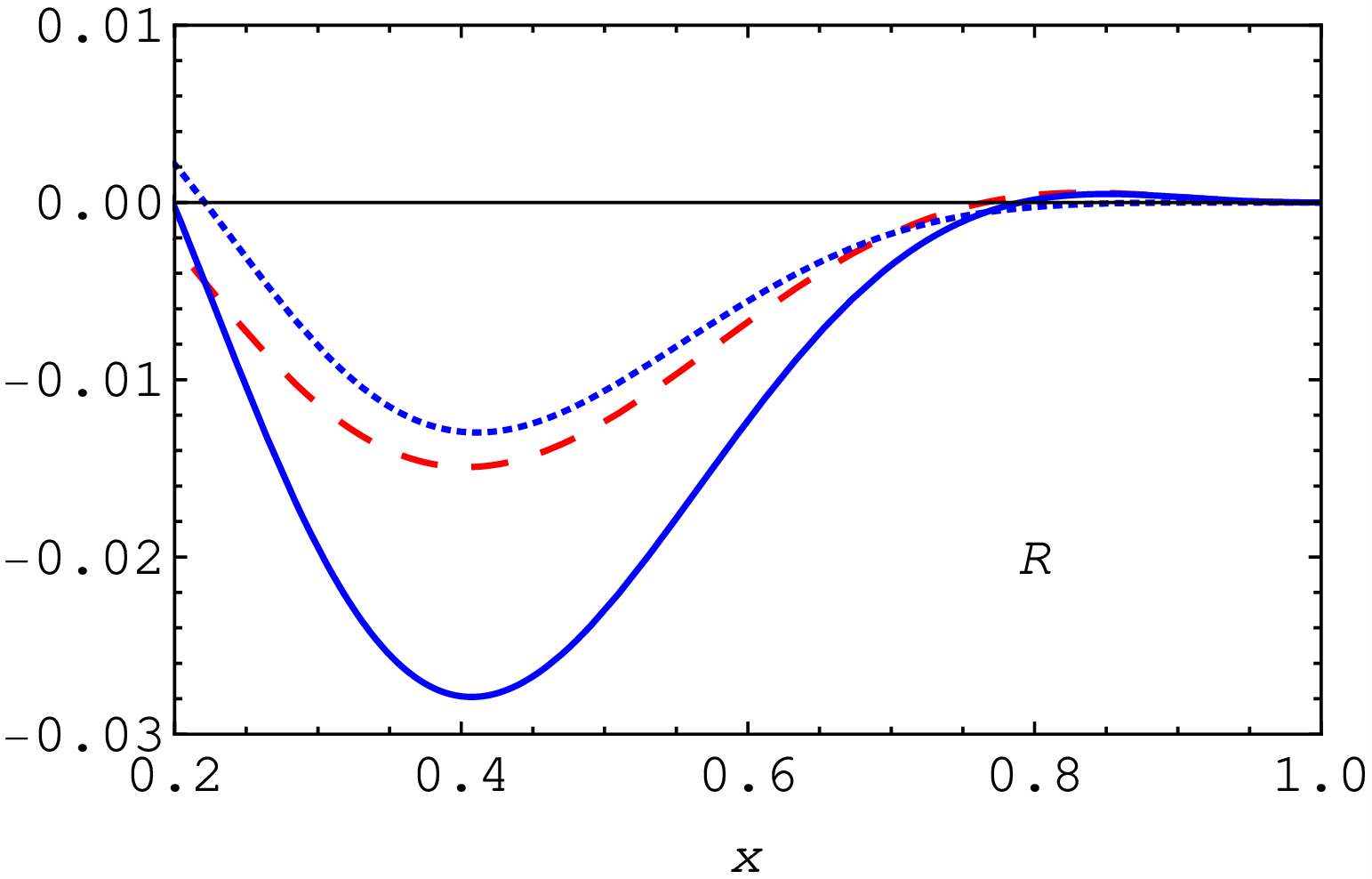}
\\[5mm]
\includegraphics[width=0.47\textwidth,clip=true]{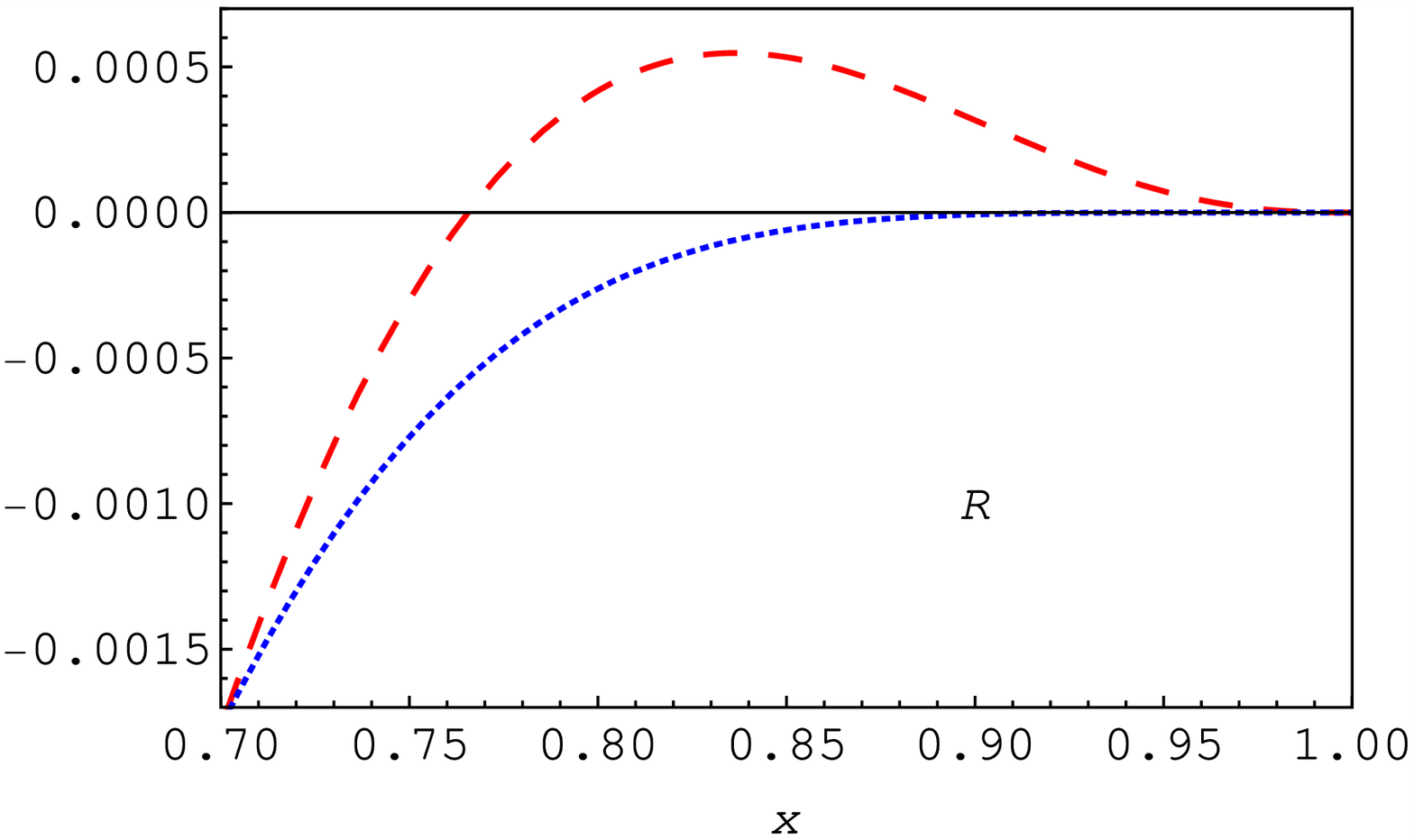}
\caption{The nucleon twist-four distribution $\widetilde{Q}^{(p)}(x)$ multiplied
by $a_3^2$ (solid line).
The dashed and dotted lines show  the contribution of three-quark and
and quark-gluon wave functions, respectively. The lower panel is a blow-up of
the high $x$ region.
}
\label{fig2}
\end{figure}

Now that we have models for the nucleon LCWFs, it is straightforward to evaluate
the matrix elements
of the four-fermion operators $\mathcal{Q}_\pm$ and constrain the momentum
fraction dependence of
the corresponding higher-twist correlator $\widetilde{\mathcal{Q}}^{(p)}$. The
distributions
${\mathcal{Q}}_{\pm}(\xi)$ defined by Eq.~(\ref{Qpm}) possess the following
support properties
\begin{align}\label{}
{\mathcal{Q}}_{\pm}(\xi)=&\theta(-\xi_1)\theta(-\xi_3)\theta(\xi_2)\theta(\xi_4)
\notag\\[2mm]
&\times \theta(1-\theta_2-\theta_4)q_{\pm}(-\xi_1,\xi_2,-\xi_3,\xi_4)\,.
\end{align}
Here the functions $q_{\pm}(\xi)$ are expressed in terms of integrals involving
the nucleon wave functions, see
Appendix~\ref{app:exp} for explicit formulas, while below we quote expressions
which correspond to the ansatzes
(\ref{Phi3}) and (\ref{WFG}). The structure of the Fock expansion corresponds to
the decomposition of the
twist-four distributions $q_\pm$ into the following three components
\begin{align}\label{qqg}
q_\pm(\xi)=q_\pm^{3q}(\xi)+q_\pm^{g_\downarrow}(\xi)
+q_\pm^{g_\uparrow}(\xi)\,.
\end{align}
Each term in this sum corresponds to the contribution of the pertinent
multi-parton component of the nucleon
wave functions, i.e., three-quark  and quark-gluon, respectively. Making use of
the results derived in the previous section,
one finds the following explicit momentum fraction dependence for the
distributions $q_\pm^{3q}$,
\begin{align}\label{q3q-exp}
q_-^{3q}(\xi)=&
c_{3q}\chi_1(\xi)\,\Big[(4-3(\xi_2+\xi_4))^2
\!+\!(5-3\xi_3)(5-3\xi_4)\Big]\,,\notag\\
q_+^{3q}(\xi)=&
c_{3q}\chi_1(\xi)\,(1+3\xi_1)(1+3\xi_2)\,,
\end{align}
where
\begin{align}\label{}
\chi_1(\xi)=&\xi_1\xi_2\xi_3\xi_4\,\frac{1-\xi_2-\xi_4}{\xi_2+\xi_4}\,,
\end{align}
and the overall normalization constant being
\begin{align}
c_{3q}=P_{3q}\,\frac{560}{87\pi^2 a_3^2}.
\end{align}
For the four-parton quark-gluon functions $q_\pm^{g_\uparrow},\,
q_\pm^{g_\downarrow} $ one gets
\begin{align}\label{qg-exp}
q_\pm^{g_\uparrow}(\xi)=&c_{g_\uparrow}^\pm \,\chi_2(\xi)\,,\notag\\[2mm]
q_\pm^{g_\downarrow}(\xi)=&c_{g_\downarrow}^\pm \,\chi_2(\xi)\,,
\end{align}
where
\begin{align}\label{}
\chi_2(\xi)=\chi_1(\xi)\,(1-\xi_2-\xi_4)^3
\end{align}
and
\begin{align}\label{}
c^+_{g_\uparrow}=&P_{g_\uparrow}\,\frac{560}{\pi^2 a_g^2}
\left[1-\frac53\frac{\lambda_3^2}{\lambda_2^2+\lambda_3^2}\right]\,,
\notag\\
c^-_{g_\uparrow}=&-P_{g_\uparrow}\,\frac{700}{\pi^2 a_g^2}
\left[1+\frac{\lambda_3}{\lambda_2^2+\lambda_3^2}\left[\frac65\lambda_2-
\frac43\lambda_3\right]\right]\,,
\notag\\
c^+_{g_\downarrow}=&P_{g_\downarrow}\,\frac{280}{\pi^2 a_g^2} \,,
\end{align}
while $c^-_{g_\downarrow}=0$.

Furthermore, making use of Eq.\ (\ref{Q2I})
one obtains after some algebra the following representation for the function
$\widetilde{\mathcal{Q}}_p(x)$, $x>0$:
\begin{widetext}
\begin{align}\label{Qexp}
\widetilde{\mathcal{Q}}_p(x)=&
-2\int_0^{1-x} d\xi\Biggl\{\frac{1}{\xi}\log \left(x/\xi\right)\,
\widehat q_+(x,\xi,\xi,x)+
\notag\\
&\frac1{x+\xi}\biggl\{
\int_0^{x+\xi}\frac{d\eta}{\eta}\biggl[\frac{x+\xi}{\eta-\xi}
\Big(\widehat q_+(x,\eta,\xi,x+\xi-\eta)-\frac{\eta}{\xi}
\widehat q_+(x,\xi,\xi,x)
\Big)
-
\widehat q_-(x,\eta,\xi,x+\xi-\eta)\biggr]+\,
\notag\\
&
\frac12\int_0^{1-x-\xi}\frac{d\eta}{\eta}
\biggl[\left[\frac{x+\xi}{\xi}+\frac{\eta}{\eta+x}\right]
\widehat q_+(\eta+x,\eta,\xi,\xi+x)
+\left[1+\frac{\eta}{\eta+x}\frac{x+\xi}{\xi}\right]
\widehat q_-(\eta+x,\eta,\xi,\xi+x)\biggr]\biggr\}
\Biggr\}\,,
\end{align}
\end{widetext}
where
\begin{align}\label{}
\widehat
q_\pm(\xi)=&\frac12\Big(1+P_{14}P_{23}\Big)\Big(1+P_{13}P_{24}\Big)q_\pm(\xi).
\end{align}
Performing the final integration is straightforward and one can obtain a closed
analytical form of
 $\widetilde{\mathcal{Q}}_p(x)$ (however, the resulting expression is quite long
and in order to
 save space it will not be displayed here).
The twist-four distribution
is displayed in the upper
 panel of Fig.~\ref{fig2}. The dashed and dotted lines correspond to its
 three-quark and quark-gluon
components, respectively. Both of them exhibit a global minimum at $x \simeq
0.4$. In the lower panel
of Fig.\ \ref{fig2}, we blow up its high-$x$ region to demonstrate the node
structure of the three-quark
contribution. As $x\to 1$ the four-parton quark-gluon component of
$\widetilde{\mathcal{Q}}_p(x)$ is
suppressed by the decay factor $(1-x)^3$ with respect to the three-quark
component. At the same time
the twist-four distribution $\widetilde{\mathcal{Q}}_p(x)$ is enhanced in
comparison with the twist-two
parton densities calculated within the same model,
$\widetilde{\mathcal{Q}}_p(x)/u_p(x)\sim
\log(1-x)$ for $x\to 1$.

\begin{figure}[htb]
\psfrag{x}[cc][cc][1.1]{$x'=2x$}
\psfrag{R}[cc][cc][1.2]{$R_1^{\rm tw-4}(x)$}
\includegraphics[width=0.47\textwidth,clip=true]{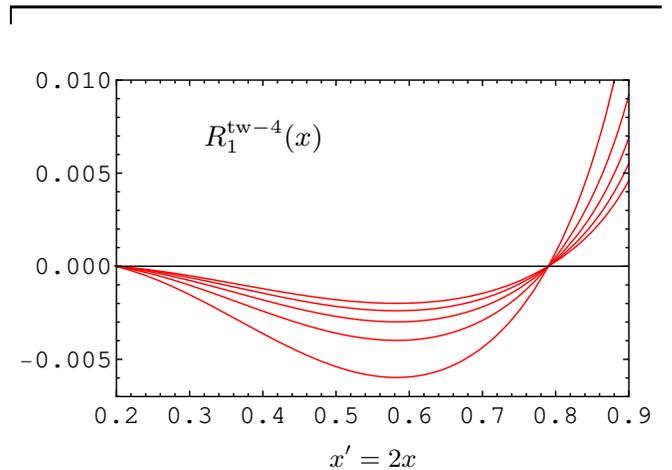}
\caption{The estimate $R_1^{\rm tw-4}$ as a function of the Bjorken $x$ for
different values of $Q^2$.
The curves from the bottom to top correspond to the values
$Q^2=4,6,8,10,12\, \text{GeV}^2$, respectively. The experimental accuracy of
SoLID is $\pm 0.005$
for $R_1^{tw-4}$ at an average $Q^2$ of 3.3 GeV$^2$ and $\langle x\rangle
=0.34$.}
\label{fig3}
\end{figure}

Our predictions for the twist-four correction $R_1^{\rm tw-4}$ to the Cahn-
Gilman formula is shown
in Fig.~\ref{fig3}. In order to make an comparison with the results of
Ref.~\cite{Mantry:2010ki}
easier,
we display $R_1^{\rm tw-4}$ for $Q^2=4,6,8,10,12\,\text{GeV}^2$. It turns out
that our prediction for
$R_1^{\rm tw-4}$ is roughly twice as large as that of Ref.~\cite{Mantry:2010ki}
with the minimum
of the function being slightly shifted towards lower $x'$ (i.e., from $x' \simeq
0.7$ to $x' \simeq 0.6$).
Note that the $x$-dependence of the twist-four contribution is much better
determined than its
normalization:
The three-quark component of the nucleon wave functions is
constrained by the existing experimental data (parton densities and nucleon form
factor,
\cite{Bolz:1996sw}), but the ansatz~(\ref{WFG}) for the quark-gluon wave
functions has to be regarded as an
exploratory estimate (see Ref.~\cite{BLMP} for a discussion). Nevertheless,
since for large $x'$ the
contribution due to the quark-gluon components of the wave functions  are
strongly suppressed,~see Fig.~\ref{fig2},
we believe  that for $x'>0.7$ our estimate for $R_1^{\rm tw-4}(x')$ should be
rather accurate. That is,
the function $R_1(x')$ has to change sign around $x'\sim 0.8$. We also checked
that our
result, once we compute its Mellin moments, are in good agreement with earlier
calculations of higher
twist corrections to the first moments of structure
functions~\cite{Fajfer:1984um,Castorina:1985uw}.

\section{Conclusion}

Parity-violating deep inelastic scattering is a process of fundamental
importance and, therefore, will be
investigated by ever more precise experiments. It is sensitive to physics beyond
the Standard
Model as well as to specific aspects of strong interaction dynamics, encoded in
higher-twist correlators.
To disentangle both, the $x$-dependence of the twist-four contribution must be
known precisely which seems
to be in reach with present day techniques. The task of determining these
higher-twist contributions
has a certain urgency in view of upcoming JLab experiment SoLID
\cite{Souder:2008zz}.
In the current study we calculated
the twist-four correction to the leading contribution $\tilde a_1$ to the parity
violating asymmetry
by determining matrix elements of light-cone four-quark operators
\cite{Bjorken:1978ry}.
We found that within the framework of light-cone wave functions, the estimate
for twist-four correlation
functions has similar features as found in a recent calculation within the MIT
bag model
\cite{Mantry:2010ki}. The size of the
correction $R_1$ is about twice as large in our calculation and the form differs
slightly, but these
differences might well reflect the present day theoretical uncertainties of such
calculations.
The size of the twist-four correction we obtain is borderline. It has to be
taken into account
to improve the sensitivity of SoLID for New Physics, but it does not seem to be
large enough for SoLID
to test our prediction. However, as mapping out the running of $\sin^2 \theta_W$
is one of the
fundamentally important experiments
we are optimistic that still more precise experiments will be performed in
future, which should then be sensitive
enough to observe the higher-twist contributions we analysed.

\section*{Acknowledgements}

The authors are grateful to V.M.~Braun for the valuable discussions. This work was
supported by DFG (grant 9209506, A.M.), BMBF (grant 06RY9191, A.M. and A.S.), RFFI (grant
09-01-93108, A.M.), and the National Science Foundation (grant No. PHY-0757394, A.B.).


\appendix
\renewcommand{\theequation}{\Alph{section}.\arabic{equation}}
\section{Light-Cone expansion}\label{app:light-cone}

In this Appendix in order to make the paper self-consistent, we spell out our
notations and conventions that
we used to perform calculations of hadronic matrix elements in the body of the
paper.

For an arbitrary four-vector $a^\mu$ we define the light-cone coordinates as
\begin{eqnarray}
&& a_+ = \frac{1}{\sqrt{2}}(a^0+a^3)\,,\qquad
 a_- = \frac{1}{\sqrt{2}}(a^0-a^3)\,,
\nonumber\\
&& a = a^1 + i a^2\,, \hspace*{1.65cm}
\bar a = a^1-i a^2\,.
\end{eqnarray}
We find it convenient to pass from four-dimensional vectors to two-dimensional
matrix notations for
all tensors. For a vector $a_\mu$ we introduce the matrix $a=a_\mu \sigma^\mu$,
where
$\sigma^\mu=(\mathbb{I},\vec{\sigma})$,
\begin{align}\label{a}
a_{\alpha\dot\alpha}=a_\mu
\sigma^\mu_{\alpha\dot\alpha}=\begin{pmatrix}\sqrt{2}a_-& -\bar a\\
-a &\sqrt{2}a_+\end{pmatrix}_{\alpha\dot\alpha} ~ .
\end{align}
In the Weil representation the Dirac $\gamma-$matrices has the form
\begin{align*}
\gamma^0= \begin{pmatrix} 0&\mathbb{I}\\\mathbb{I}&0\end{pmatrix}\,,&&
\gamma^i= \begin{pmatrix} 0&\sigma^i\\-\sigma^i&0\end{pmatrix}\,,
&&
\gamma^5 =
\begin{pmatrix} -\mathbb{I}& 0 \\ 0& \mathbb{I}\end{pmatrix}\,,
\end{align*}
with $\gamma^5 = i\gamma^0\gamma^1\gamma^2\gamma^3$.
In the two-component notation the Dirac spinors read
\begin{align}
   q =
 \begin{pmatrix}q_\downarrow\\ q_\uparrow \end{pmatrix}
   \,,\quad
\bar q = q^\dagger\gamma^0 =
(\bar q_\downarrow,\bar q_\uparrow)\,,
\end{align}
where $q_{\uparrow(\downarrow)}=\dfrac12(1\pm\gamma_5)q$ are components with
positive/negative helicity,
respectively. The two independent light-like vectors
\begin{align}
n^\mu = \frac{1}{\sqrt{2}}(1,0,0,-1)\,, \qquad \tilde n^\mu =
\frac{1}{\sqrt{2}}(1,0,0,1)\,,
\end{align}
$n^2=\tilde n^2=0$, $n \cdot \tilde n=1$ can be parameterized in terms of two
auxiliary Weil spinors:
\begin{equation}
n_{\alpha\dot\alpha}=\lambda_\alpha\bar \lambda_{\dot\alpha}\,,\qquad
\tilde n_{\alpha\dot\alpha}=\mu_\alpha\bar \mu_{\dot\alpha}\,,
\end{equation}
which read explicitly
\begin{eqnarray}\label{lambdamu}
\lambda_\alpha=2^{1/4}\begin{pmatrix}-1 \\0\end{pmatrix},&&
\mu_\alpha=2^{1/4}\begin{pmatrix}0 \\1\end{pmatrix},
\nonumber\\
\bar\lambda_{\dot\alpha}=2^{1/4}\begin{pmatrix}-1 \\ 0 \end{pmatrix},&&
\bar\mu_{\dot\alpha}=2^{1/4}\begin{pmatrix}0  \\ 1\end{pmatrix}.
\end{eqnarray}
The following rules allow to raise and lower spinor indices
\begin{align*}
\lambda^\alpha=\epsilon^{\alpha\beta}\lambda_\beta\,,&&
\lambda_\alpha=\lambda^\beta\epsilon_{\beta\alpha}\,,&&
\bar\lambda^{\dot\alpha}=\bar
\lambda_{\dot\beta}\epsilon^{\dot\beta\dot\alpha}\!\!, &&
\bar \lambda_{\dot\alpha}=\epsilon_{\dot\alpha\dot\beta}\bar
\lambda^{\dot\beta}\!,
\end{align*}
with the antisymmetric Levi-Civita tensor having only the following
nonzero components
$$
\epsilon_{12}=\epsilon^{12}=
-\epsilon_{\dot 1\dot 2}=-\epsilon^{\dot 1\dot 2}=1\,.
$$
The auxiliary spinors $\lambda$ and $\mu$ are normalized as
\begin{align}
 (\mu\lambda) = \mu^\alpha\lambda_\alpha = -(\lambda\mu) = - \sqrt{2}\,,
\nonumber\\
 (\bar\mu\bar\lambda) =
  \bar\mu_{\dot\alpha}\bar\lambda^{\dot\alpha} = -(\bar\lambda\bar \mu) = +
\sqrt{2}\,
\end{align}
and are used to project out the ``plus" and ``minus" components of the fields.
For fermions,
we define
\begin{eqnarray}
\psi_+=\lambda^\alpha \psi_\alpha, &&
\psi_-=\mu^\alpha \psi_\alpha,
\nonumber\\
\bar \chi_+= \bar\chi_{\dot\alpha}\bar\lambda^{\dot\alpha},&&
\bar \chi_-=  \bar\chi_{\dot\alpha}\bar\mu^{\dot\alpha} \,.
\end{eqnarray}
In the same fashion the light-cone decomposition of a vector (e.g., gluon) field
takes the form
\begin{align*}
  A_{\alpha\dot\alpha} &=  A_- \,\lambda_\alpha\bar\lambda_{\dot\alpha}
                   + A_+ \,\mu_\alpha\bar\mu_{\dot\alpha}
                   + \frac{\bar A}{\sqrt{2}} \, \lambda_\alpha
\bar\mu_{\dot\alpha}
                   + \frac{A}{\sqrt{2}}\, \mu_\alpha \bar\lambda_{\dot\alpha}\,.
\end{align*}
The ``plus" spinor fields $\psi_+,\bar\chi_+$ and transverse gluon fields
$A,\bar A$
are assumed to be the dynamical degrees of freedom in the light-cone
quantization framework.
While the  ``minus" fields $\psi_-,\bar\chi_-, A_-$ can be expressed in terms of
these
with the help of equations of motion. Finally, we use the gauge
$A_+=0$.

The {\em good} components of the quark field have the following canonical
expansion
\begin{align}\label{qupdown}
q_{\downarrow+}(x)=&
\int\! \frac{dp_+}{\sqrt{2p_+}}  \frac{d^2 \bit{p}_\perp}{(2\pi)^3} \theta(p_+)
\biggl[ e^{-ip \cdot x} b_\downarrow(p)+e^{+ip \cdot x}
d_\uparrow^\dagger(p)\biggr],
\notag\\
q_{\uparrow+}(x)=&\int\! \frac{dp_+}{\sqrt{2p_+}}  \frac{d^2
\bit{p}_\perp}{(2\pi)^3} \theta(p_+)
\biggl[ e^{-ip \cdot x} b_\uparrow(p)+e^{+ip \cdot x}
d^\dagger_\downarrow(p)\biggr],
\end{align}
in terms of the annihilation operators of quark and antiquark of positive
(negative) helicity
$b_{\uparrow (\downarrow)},d_{\uparrow (\downarrow)}$, respectively. They obey
the
standard anticommutation relations
\begin{multline}
\label{bbdd}
\{b_{\lambda}(p),b_{\lambda'}^\dagger(p')\}=\{d_{\lambda}(p),d^\dagger_{\lambda'
}(p')\}=
\\
=2p_+ (2\pi)^3\delta_{\lambda,\lambda'}\delta(p_+-p'_+)\delta^{(2)}(
\bit{p}_\perp- \bit{p}'_\perp)\,.
\end{multline}
Similarly the expansion for the dynamical transversely polarized gluon fields
$A$ and $\bar A$ reads
\begin{eqnarray}
\bar A(x)\!\!\!&=&\!\!\!\sqrt{2}\!\int\! \frac{dk_+}{2k_+}  \frac{d^2
\bit{k}_\perp}{(2\pi)^3} \theta(k_+)
\biggl[ e^{-ik \cdot x} a_\uparrow (k)+e^{+ik \cdot x} a_\downarrow^\dagger(k)
\biggr],
\nonumber\\
A(x)\!\!\!&=&\!\!\!\sqrt{2}\!\int\! \frac{dk_+}{2k_+}  \frac{d^2
\bit{k}_\perp}{(2\pi)^3} \theta(k_+)
\biggl[ e^{-ik \cdot x} a_\downarrow (k)+e^{+ik \cdot x}
a_\uparrow^\dagger(k)\biggr].
\nonumber\\
\label{a+a}
\end{eqnarray}
Here and below $A=\sum_a t^a A^a$ are matrices in the fundamental representation
of $SU(3)$
and $t^a$ are the usual generators, normalized as ${\rm tr}(t^a t^b)=\dfrac12
\delta^{ab}$.
The creation and annihilation operators obey the commutation relation
\begin{eqnarray}
\lefteqn{\Big[a^b_\lambda(p),(a^{b'}_{\lambda'}(p'))^\dagger\Big]=}
\\&=&
2p_+ (2\pi)^3\delta_{\lambda,\lambda'}\delta^{bb'}\delta(p_+-
p'_+)\delta^{(2)}(\bit{p}_\perp-\bit{p}'_\perp)\,.
\nonumber
\end{eqnarray}
As mentioned above, {\em bad} (i.e., ``minus'') components can be expressed in
terms of the dynamical
fields using QCD equations of motion.

\section{Distributions $q_{\pm}$}\label{app:exp}
%
As discussed in the main text in Sect.\ \ref{sect:results}, we represent  the
twist-four distributions $q_\pm (\xi)$ as shown in
Eq.~(\ref{qqg}). We remind here that the arguments $\xi = (\xi_1, \xi_2, \xi_3,
\xi_4)$ are subject to
the constraints, $0\leq \xi_i\leq 1$ and $\xi_1+\xi_3=\xi_2+\xi_4$. A
straightforward calculation of its components
$q_\pm^{3q}(\xi),$ $q_\pm^{g_\downarrow}(\xi),$ $q_\pm^{g_\uparrow}(\xi)$,
arising from three- and
four-parton Fock states of the nucleon, yields the following expressions in
terms of the LCWFs introduced in the
main text,
\begin{widetext}
\begin{align}\label{}
q_+^{3q}(\xi)=&-\frac{4}{9}\frac{(\pi a_3 f_N)^2}{(\xi_2+\xi_4)(1-\xi_2-\xi_4)}
\phi(\xi_1,1-\xi_1-\xi_3,\xi_3)\phi(\xi_2,1-\xi_2-\xi_4,\xi_4)\,,
\notag\\
q_-^{3q}(\xi)=&-\frac{4}{9}\frac{(\pi a_3 f_N)^2}{(\xi_2+\xi_4)(1-\xi_2-
\xi_4)}\biggl\{
\phi(1-\xi_1-\xi_3,\xi_1,\xi_3)\phi(1-\xi_2-\xi_4,\xi_2,\xi_4)
\notag\\
&
+\Big(\phi(\xi_1,\xi_3,1-\xi_1-\xi_3)+\phi(1-\xi_1-\xi_3,\xi_3,\xi_1)\Big)
\Big(\phi(\xi_2,\xi_4,1-\xi_2-\xi_4)+\phi(1-\xi_2-\xi_4,\xi_4,\xi_2)\Big)
\biggr\}\,,
\end{align}
where $\phi$ is given by Eq.~(\ref{phi3}). Next, we got that $q_-
^{g_\downarrow}(\xi)=0$
and
\begin{align}\label{}
q_+^{g_\downarrow}(\xi)=&\frac{32(8\pi^2 a_g^2)^2}{3(\xi_2+\xi_4)}
\int_0^1 \frac{dx_2}{x_2} \frac{dx_4}{x_4^2}\delta(1-\xi_2-\xi_4-x_2-x_4)
\Biggl\{
\phi_g(\xi_1,x_2,\xi_3,x_4)\Big[\phi_g(x_2,\xi_2,\xi_4,x_4)+\frac14\phi_g(\xi_2,
x_2,\xi_4,x_4)\Big]
\notag\\
&
+\phi_g(x_2,\xi_1,\xi_3,x_4)\Big[\phi_g(\xi_2,x_2,\xi_4,x_4)
-2\phi_g(x_2,\xi_2,\xi_4,x_4)\Big]
\Biggr\}\,.
\end{align}
Finally,
\begin{align}\label{}
q_+^{g_\uparrow}(\xi)=&\frac{8(8\pi^2 a_g^2)^2}{3(\xi_2+\xi_4)}
\int_0^1 \frac{dx_2}{x_2} \frac{dx_4}{x_4^2}\delta(1-\xi_2-\xi_4-x_2-x_4)
\Biggl\{
\psi_g^{(1)}(\xi_1,x_2,\xi_3,x_4)\Big[\psi^{(1)}_g(\xi_2,x_2,\xi_4,x_4)+
5\psi^{(2)}_g(\xi_2,\xi_4,x_2,x_4)\Big]
\notag\\
&
+\psi^{(2)}_g(\xi_1,\xi_3,x_2,x_4)
\Big[\psi^{(2)}_g(\xi_2,\xi_4,x_2,x_4)+5\psi^{(1)}_g(\xi_2,x_2,\xi_4,x_4)\Big]
\Biggl\}\,,
\notag\\
q_-^{g_\uparrow}(\xi)=&\frac{32(8\pi^2 a_g^2)^2}{3(\xi_2+\xi_4)}
\int_0^1 \frac{dx_2}{x_2} \frac{dx_4}{x_4^2}\delta(1-\xi_2-\xi_4-x_2-x_4)
\Biggl\{
\psi_g(\xi_1,x_2,\xi_3,x_4)\Big[\psi_g(x_2,\xi_2,\xi_4,x_4)+\frac14\psi_g(\xi_2,
x_2,\xi_4,x_4)\Big]
\notag\\
&
+\psi_g(x_2,\xi_1,\xi_3,x_4)\Big[\psi_g(\xi_2,x_2,\xi_4,x_4)
-2\psi_g(x_2,\xi_2,\xi_4,x_4)\Big]
-\Big[\psi^{(1)}_g(x_2,\xi_2,\xi_4,x_4)-\frac14
\psi^{(2)}_g(x_2,\xi_4,\xi_2,x_4)\Big]
\notag\\
&\times\psi_g^{(2)}(x_2,\xi_3,\xi_1,x_4)-
\psi_g^{(1)}(x_2,\xi_1,\xi_3,x_4)\Big[\psi^{(2)}_g(x_2,\xi_4,\xi_2,x_4)
+2\psi^{(1)}_g(x_2,\xi_2,\xi_4,x_4)\Big]
\Biggr\}\,,
\end{align}
\end{widetext}
where
\begin{align*}\label{}
\psi_g(x_1,x_2,x_3,x_4)=
\psi_g^{(1)}(x_1,x_2,x_3,x_4)-\psi_g^{(2)}(x_3,x_1,x_2,x_4)\,.
\end{align*}
%


\end{document}